\documentclass{article}
\usepackage{framed}
\usepackage{geometry}
\usepackage{amssymb}
\usepackage{amsmath}
\usepackage{color,xcolor}
\usepackage[colorlinks,linkcolor=blue,anchorcolor=blue,citecolor=green]{hyperref}
\usepackage{booktabs}
\usepackage{hyperref}

\usepackage{graphicx}
\usepackage[labelfont=bf]{caption}
\usepackage{titlesec}
\usepackage{titling}
\usepackage{url}

\begin{document}

\urlstyle{same}



%
%


\begin{center}
    \Large \textsf{\textbf{BubbleNet: Inferring micro-bubble dynamics with semi-physics-informed deep learning}}
\end{center}

\begin{center}
    \large Hanfeng Zhai$^{a,b}$, Quan Zhou$^{a}$ and Guohui Hu$^{a,\star} $
\end{center}

\begin{center}
    $^{a}$\textit{Shanghai Institute of Applied Mathematics and Mechanics},\\
    \textit{School of Mechanics and Engineering Science},\\
    \textit{Shanghai Key Laboratory of Mechanics in Energy Engineering},\\
    \textit{ Shanghai University, Shanghai 200072, PR China }\\ \vspace{5pt}
    $^{b}$\textit{Sibley School of Mechanical and Aerospace Engineering,\\ Cornell University, Ithaca 14850, NY, USA}
\end{center}

\begin{center}
    $^\star$Corresponding author: \textsf{ghhu@staff.shu.edu.cn}
\end{center}

\date{\today}
\begin{abstract}
Utilizing physical information to improve the performance of the conventional neural networks is becoming a promising research direction in scientific computing recently. For multiphase flow, it would require significant computation resources for neural network training due to the large gradients near the interface between the two fluids. Based on the idea of the physics-informed neural networks (PINNs), a deep learning framework \textsf{BubbleNet} is proposed to overcome this difficulty in the present study. The deep neural networks (DNN) with sub-nets is adopted to predict physics fields, with the semi-physics-informed part encoding the continuum equation and the pressure Poisson equation $\mathcal{P}$ for supervision, and the time discretized normalizer (TDN) is also used to normalize field data per time step before training. Two bubbly flows, i.e., single bubble flow and multiple bubble flow in a microchannel, are considered to test the algorithm. The conventional CFD software is applied to obtain the training data set. The traditional DNN and the \textsf{BubbleNet}(s) are utilized to train the neural network and predict the flow fields for the two bubbly flows. Results indicate the \textsf{BubbleNet} framework is of ability to successfully predict the physics fields, and the inclusion of the continuum equation significantly improves the performance of deep NN. The introduction of the Poisson equation also has slightly positive effects on the prediction results. The results suggest that constructing semi-PINN by flexibly considering the physical information into neural networks will be helpful in the learning of complex flow problems.
\end{abstract}






\section{Introduction\label{intro}}

Machine learning (ML) has achieved tremendous success in the last decade due to the availability of big data and computer resources. ML is the study of algorithms that allows computer programs to automatically improve their performance through experience \cite{Mitchell1997}. AlphaGo burst the public's interest by showing the huge potential of machine learning and artificial intelligence \cite{Silver2016, Silver2017}. The ML techniques 
is becoming a promising research method in diverse scientific fields, specifically in genomics \cite{aFold, geno}, public health \cite{med2, ph1, ph2, cov1, cov2}, and medicine \cite{med1, med3, ct, ct2}. 

Deep neural networks (DNNs), as one of the most prominent tools of ML, have been adopted to tackle various physics problems including turbulence \cite{turbulence}, flow control \cite{flowcontrol}, heat transfer \cite{heattransfer}, and combustion \cite{combustion}. These deep learning applications have grown drastically in recent years, mainly on learning physical equations and inferring dynamics. Numerous frameworks has henceforth been proposed, such as SINDy \cite{sindy} and PDE-FIND \cite{pdefind}, using sparse regression to identify the governing equations for nonlinear dynamic systems; Graph Kernel Network \cite{graphnet}, Fourier Neural Operator \cite{fno} and MeshfreeFlowNet \cite{meshnet}, using convolutional neural networks to learn image mapping  physics fields; Deep potential \cite{deeppot}, DeePMD \cite{deepmd} and DeePCG \cite{deepcg}, using deep neural nets to map the molecular potentials at the microscale. In 2018, Raissi et al. \cite{pinn, pinn1, pinn2} proposed a deep learning framework called physics-informed neural networks (PINNs) for identifying and inferring dynamics of physical systems governed by partial differential equations (PDEs). The strategy of PINN can be simplified as encoding governing PDEs into the loss function as a soft physics constraint, namely the 'physics-informed' part. Based on PINN, Lu et al. proposed DeepXDE \cite{deepxde} and DeepONet \cite{deeponet}, the refined versions of PINNs, for learning and inferring nonlinear operators of PDEs, which were later applied to electroconvection multiphysics \cite{deeponeta} and hypersonics \cite{deeponetb}.

The PINN series has been developed to solve numerous problems, including Fractional PINNs for predicting fractional PDEs; Conservative PINNs for nonlinear conservation laws \cite{cpinn}; Extended PINNs, a PINN approach for space-time domain decomposition \cite{xpinn}; Parareal PINNs, a PINN solver decomposing a long-time problem into many independent short-time problems supervised by an inexpensive/fast coarse-grained (CG) solver \cite{ppinn}. PINNs has achieved great success for predicting laminar flows \cite{pinn-laminar}, high speed flows \cite{pinn-high}, heat transfer \cite{pinn-heat} and turbulence \cite{pinn-turbulence}. Lin et al. \cite{deeponetbubble} have used the prementioned DeepONet to predict bubble growth dynamics described by Rayleigh–Plesset (R–P) equation.

Despite the significant development of Informed Machine Learning \cite{informedml} architecture, the consideration of physical equations in loss function usually requires high order differentiation of physical quantities. Specifically, in two-phase flows, the phase value at the interface between different fluids exhibits a drastic variation from 0 to 1, making the calculation of the gradient highly difficult. Therefore, high-resolution training data would be a prior condition for the success of the algorithm, especially for variables with high gradients. This will greatly increase the amount of deep learning computation. Furthermore, high-resolution data can hardly be obtained in many experiments. 

In the present study, we provide an engineering-orientated idea to overcome this difficulty. Bubbly flows are considered to test the algorithm because it is a classic fluid mechanics problem with a high gradient of density. It has widely been applied in biomedical engineering, such as blood-brain barrier \cite{bbb1, bbb2, bbb3} and drug delivery \cite{bmed1, bmed2}. The bubble pinch-off effect confined in a microchannel is one of the most studied phenomena in fluid mechanics \cite{bp1, bp2, bp3, bp4, bp5}, depicting deformation and movement of single bubble dynamics. The flow with multiple bubbles displays complexity due to the interactions between bubbles \cite{bsys1, bsys2}.

Our work is inspired by the ideas of PINNs \cite{pinn}, and the subsequent work using DeepONet to infer bubble dynamics by Lin et al. \cite{deeponetbubble}. Instead of applying the full hydrodynamic equations to supervise the training process, we aim at getting satisfactory results based on partial physical information. Specifically, we inserted only the fluid continuity condition (divergence free of velocity) and the pressure Poisson equation (denoted by $\mathcal{P}$) into the loss function, which can be described as a neural network with semi-physical information. A new deep learning architecture, called \textsf{BubbleNet}, is proposed for this purpose. 


The \textsf{BubbleNet}(s) with/without Poisson equation are considered and trained in the present study. The advantages of the algorithms lie in: (1) To save computer resources, the gradient computation from automatic differentiation \cite{autodiff} of the phase function is avoided to some extent. (2) Both physical information and conventional NN losses are encoded in the semi-physics-informed network, allowing the framework to take advantage of deep learning and physical equations during training. (3) In 2D or axisymmetric flow, the velocities are inferred from a latent function (stream function), saving computation resources compared with inferring two components of velocities separately. Another point that differs from the traditional approach is, the time discretized normalizer (TDN) is utilized for normalizing variables per time step, to capture physics information more accurately for the NN training. 


The paper is arranged as follows: in Section \ref{bubble}, to obtain the training data, the conventional computational fluid dynamics (CFD) software is utilized to simulate numerically two cases of the bubble flow, i.e., the single bubble flow and multiple bubbles flow through a microchannel. Then the numerical results are briefly discussed. In Section \ref{nn} we introduce the traditional DNNs and the \textsf{BubbleNet} algorithms, respectively. Both networks are trained based on the numerical results. We hence obtain the predictions of physical quantities in bubble flows and analyze the absolute errors from machine learning in Section \ref{results}. Finally, some conclusions of this study are drawn in Section \ref{conclusion}. 

\section{Numerical implementation of bubble flows\label{bubble}}

\subsection{Problem formulation}

Bubble flows are commonly encountered in numerous
biological applications. Two phase flows (air and water) are governed by the Navier-Stokes equation:
\begin{equation}\rho \left( \frac{\partial \mathbf{u}}{ \partial t} + (\mathbf{u} \cdot \nabla) \mathbf{u}\right) =  - \nabla p + \mu \nabla^2 \mathbf{u}\label{ns}\end{equation}
where $\rho$ is the density of the fluid, $\mathbf{u} = (u,\ v)$ is 2D velocity vector, $p$ is the fluid pressure and $\mu$ is the dynamic viscosity. 

In the present study, the level set algorithm is applied to bubble flows in the microchannel, to obtain the data set for machine learning training. In the level set method, the interface between air and water is represented by a certain level set or isocontours of a globally defined function, e.g., the level set function $\phi = \phi(x,\ y,\ t)$ in 2D spaces \cite{comsol}. In our system, $\phi$ is a smooth step function that equals zero for water and one for air. Across the interface, there is a smooth transition from zero to one. Thus, the interface is defined by 0.5 in level set $\phi$.

The level set phase function $\phi$ takes the form: \begin{equation}\frac{\partial \phi}{ \partial t} + \nabla \phi \cdot \mathbf{u} = F\label{lsns}\end{equation}
where $F$ includes terms with higher-order derivatives of $\phi$, designed to keep the interface compact. 

Bubble flows involve interactions between two fluids with different physical properties. Here, we set $\rho_l$ as the water (liquid) density and $\rho_g$ as the air (gas) density; $\mu_l$ as the water viscosity and $\mu_g$ as the air viscosity. The density and viscosity in the flow can be connected through the level set function, as follows:
\begin{equation}
\begin{aligned}
\rho = \rho_l + \phi (\rho_g - \rho_l)\\
\mu = \mu_l + \phi (\mu_g - \mu_l)
\end{aligned}
\label{lg}
\end{equation}

To simulate the interfaces between liquid and gas, equation (\ref{lsns}) can be rewritten in the following form:\begin{equation} {\frac{\partial \phi }{\partial t}} + {\mathbf{u}} \cdot \nabla \phi = \gamma \nabla \cdot \left( \epsilon_{ls} \nabla \phi - \phi (1-\phi) \frac{\nabla \phi }{ |\nabla \phi |}\right)\label{levelseteq}\end{equation}
where the terms on the left-hand side describe the motion of the interface, while those on the right-hand side are necessary for numerical stability. $\gamma$ is the reinitialization parameter, which determines the amount of reinitialization or stabilization of the level set function, equals to 1 in our cases. $\epsilon_{ls}$ is the parameter controlling the interface thickness, equals to the mesh largest size \cite{comsol}, as shown in Table \ref{tabeps}.  

The continuum equation for fluid is written as:\[\frac{\partial \rho }{ \partial t} + \nabla \cdot (\rho \mathbf{u}) = 0\]
where for incompressible fluid, such an equation can be simplified to $\partial_y u + \partial_x v = 0$, i.e., the divergence free condition of velocity, $\nabla \cdot \mathbf{u} = 0$. As we will discuss later, our goal is to insert such a condition into the neural networks (NN) for better training and predictions. 

The geometric properties of interfaces can be described by unit normal to the interface $\mathbf{n} = \left.{\nabla\phi / |\nabla\phi|}\right. _{\phi=0.5}$. The curvature of the level set phase function $\phi$ can be calculated as $\kappa_{ls} = - \left.\nabla \cdot \mathbf{n}\right|_{\phi = 0.5}$. To simulate the bubble flows numerically, equation (\ref{levelseteq}) is discretized and solved for given boundary conditions (BCs), initial conditions (ICs), and a specific space-time domain with meshing for discretizations.

\subsection{Numerical setup}

\begin{figure}
    \centering
    \includegraphics[scale=0.22]{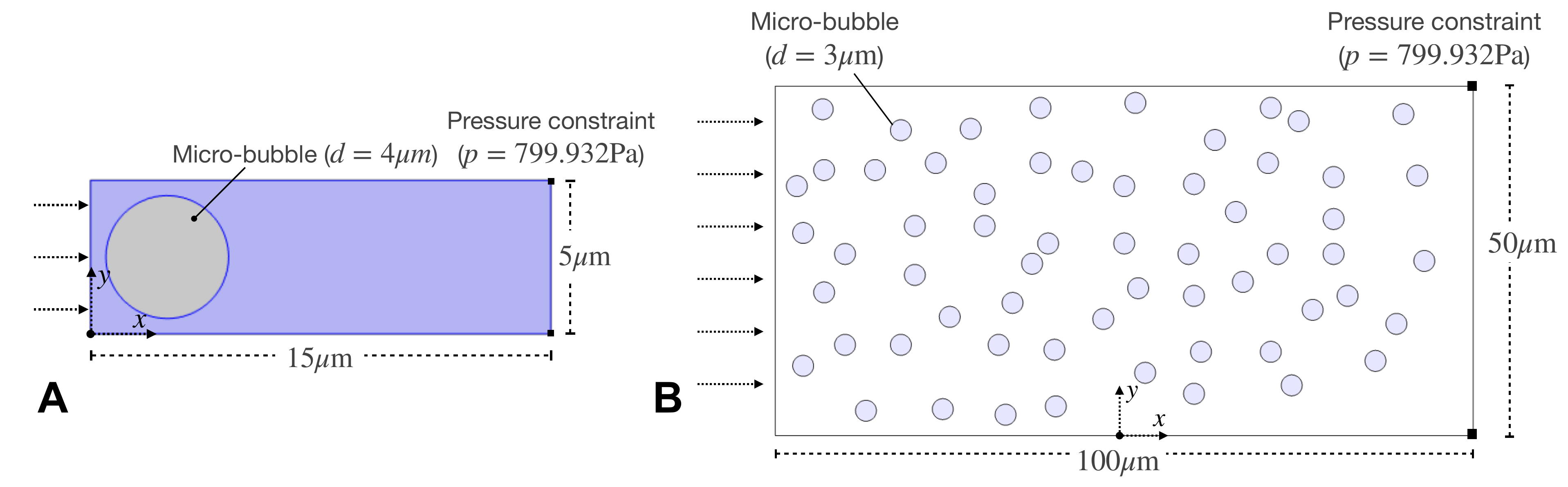}
    \caption{Diagram of the two bubble flow cases. \textbf{A.} Model for the single bubble flow case. The grey area indicates the initial position of the bubble, with diameter $d = 4 \mu \rm m$, and the blue area indicates the water. The bubble is constrained in a 2D microchannel with length equals $15 \mu\rm m$ and height of $5 \mu\rm m$. The pressure difference $\Delta p = 10\rm Pa$ is imposed between the two sides of the microchannel. The initial condition is given as a constant pressure of $799.932\rm Pa$. The coordinate is centered at the left bottom point of the microchannel. \textbf{B.} Model for the multiple bubbles flow case. The blue circles indicate 60 bubbles with each diameter $d = 4 \mu \rm m$ constrained in a 2D microchannel of length equals $100\mu\rm m$ and height $50\mu\rm m$, with surrounding fluid of water. The BCs and ICs are same as the single bubble flow case. The coordinates' origin is located at the the mid-bottom point in \textbf{B}.}
    \label{model}
\end{figure}

Two cases are considered for investigations: $(1)$ single bubble flow and $(2)$ multiple bubbles flow, confined in a microchannel. For the single bubble case, the initial diameter of the bubble is set to be $d = 4 \mu \rm m$, and the microchannel has length of $15 \mu \rm m$ and width of $5 \mu \rm m$. The pressure difference $\Delta p = 10 \rm Pa$ is imposed in axial direction to drive the flow, and pressure at the end of the channel is kept as constant pressure $p_0 = 799.932 \rm Pa$ (6mmHg), corresponding to the pressure of interstitial fluid in human brain \cite{interstitial1, interstitial2} and in lymph flow \cite{interstitial3}. The initial conditions (ICs) is set as the pressure $p_0$, with room temperature as $293.15$K, shown as in Figure \ref{model}\textbf{A}. This kind of numerical setup has been designed to simulate the bubble transportation in brain vessels for the investigation of the blood-brain barrier \cite{bbb1, bbb2, bbb3}. Our model is inspired by the work of Miao et al. \cite{bbb1}, where a single bubble is confined in a microchannel with both diameters $\sim 5 \mu$m.

For the multiple bubbles flow, 60 micro-bubbles, each micro-bubble of diameter $3 \mu \rm m$, are randomly distributed in a 2D microchannel with a length of $100 \mu \rm m$ and width of $50 \mu \rm m$, as shown in Figure \ref{model}\textbf{B}. The BCs and ICs are the same as the single bubble case. The moving mesh is adopted for space field discretization and computations. The single bubble flow case generates 24182 meshes and the multiple bubbles flow generates 75302 meshes initially. For the single bubble case, the simulation is run for $5000 \mu \rm s$, while for the multiple bubbles case, the simulation is run for $3000 \mu \rm s$.  $Re = \rho U L / \mu$, where $L$ is the characteristic length, equals to the microchannel's width in our cases. $\mu$ is the dynamic viscosity of the bubbly flow. $U = -{R^2 / 3\mu}\cdot dp/dx$ is the average velocity of 2D Poiseuille flow. $R=L/2$ is the half width of the channel. The Reynolds number $Re$ is approximately 0.007 for the single bubble case and 0.010 for the multiple bubbles case. 

To produce training data set of the bubble flows in the microchannel, the computational fluid dynamics (CFD) technique is utilized for 2D bubbly flow simulations, adopting the time-dependent level set algorithms for modeling bubbles, using COMSOL Multiphysics\textregistered. 


\begin{table*}[!b]
    \centering
    \begin{tabular}{ccc}

    \toprule
    $\epsilon_{ls}$& Dense meshing & Coarse meshing \\
    \midrule 
    Single bubble flow & 0.430 & 4.382 \\
    Multiple bubbles flow & 0.083 & 0.299\\
    \bottomrule
    \end{tabular}
    \caption{The level set parameter $\epsilon_{ls}$ value for both the two bubbly flow cases with dense and coarse meshing, respectively.}
    \label{tabeps}
\end{table*}

In subsequent sections, the numerical results obtained in CFD simulation are used for training the NNs. However, as we claimed before, high resolution data will lead to a huge demand of computer resources in NN training or maybe not available in experiments. For this consideration, we only use a "coarsened" (approximately $1/10$) data set for training NN. Therefore, we also perform CFD simulations for the two bubble flows with coarse meshing (also approximately $1/10$ of the dense meshing) for comparison, i.e., 2419 meshes for the single bubble flow and 7833 meshes for the multiple bubbles flow. In solving equation (\ref{levelseteq}), $\epsilon_{ls}$ is related with the meshing size. Hence $\epsilon_{ls}$ is listed for both the four cases in Table \ref{tabeps} for reference.

\subsection{CFD results}

\begin{figure}
    \centering
    \includegraphics[scale=0.18]{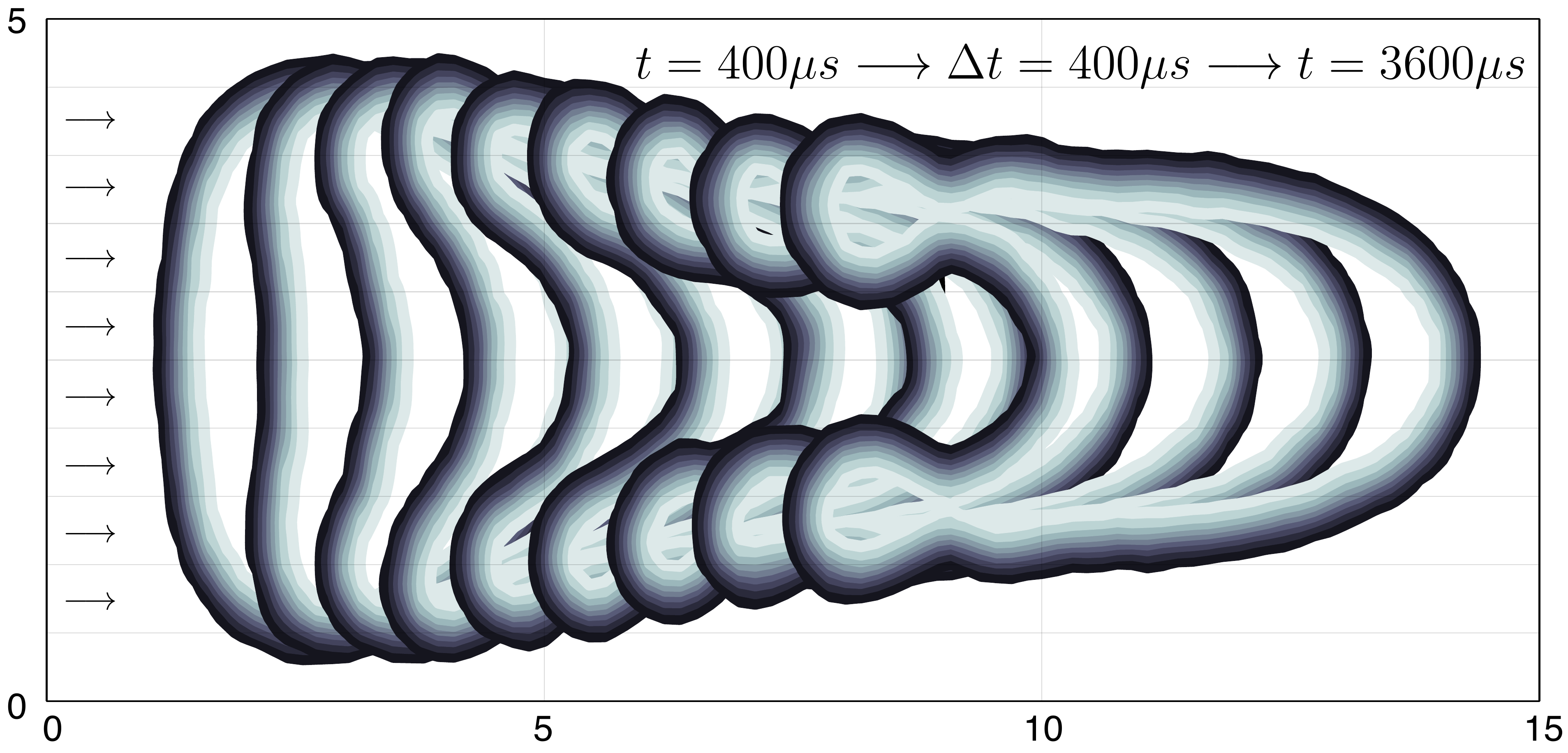}
    \caption{Simulation results depicting the bubble movement (liquid-gas interface) of the single bubble flow. The bubble deformation is illustrated at 9 different time step, starting from $400 \mu \rm s$ to $3600 \mu \rm s$, with time intervals of $400 \mu \rm s$.}
    \label{singlemove}
\end{figure}

The single bubble motion in microchannel is depicted in Figure \ref{singlemove}, where the snapshots of bubbles is illustrated at 9 different time steps, starting from $t=400 \mu \rm s$ to $t=3600 \mu \rm s$ with interval $\Delta t=400 \mu \rm s$. It shows that the front side of the bubble flows forwards in a parabolic shape since the flow velocity is relatively higher in the middle of the channel, while the rear side of the bubble is significantly stretched due to viscous shear. The configuration of the bubble is similar to the deformation of red blood cells traveling in microchannel reported by Tomaiuolo et al. \cite{rbc}.

\begin{figure}
    \centering
    \includegraphics[scale=0.7]{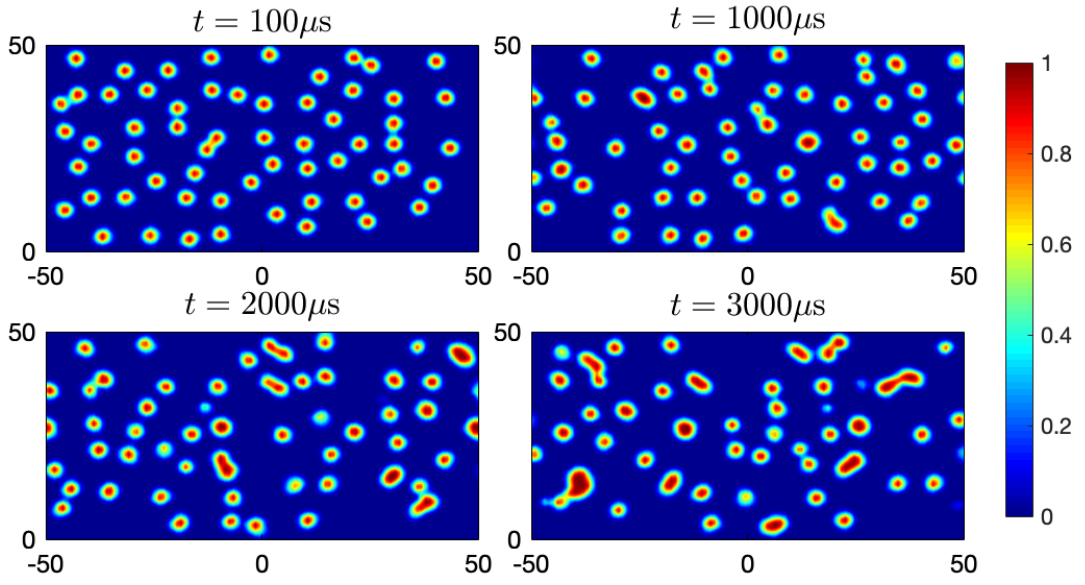}
    \caption{Snapshots of the multiple bubbles flow at different time steps, starting from $100\mu \rm s$ to $3000\mu$s. The two-phase flow is depicted by the phase function $\phi$, where $\phi = 0$ for water and $\phi = 1$ for air in the figures.}
    \label{sysmove}
\end{figure}

The multiple bubbles motion indicates bubbles tend to collide and ruptured with each other, as shown in Figure \ref{sysmove}. The collision of two 'daughter' bubbles, which is demonstrated by the contact of bubbles' outer interface (light blue part), results in the formation of bigger bubbles, as can be observed in every consecutive subfigure. Such results are consistent with the numerical study of bubble behaviors by Li et al. \cite{bubbles}.

\begin{figure}
    \centering
    \includegraphics[scale=0.26]{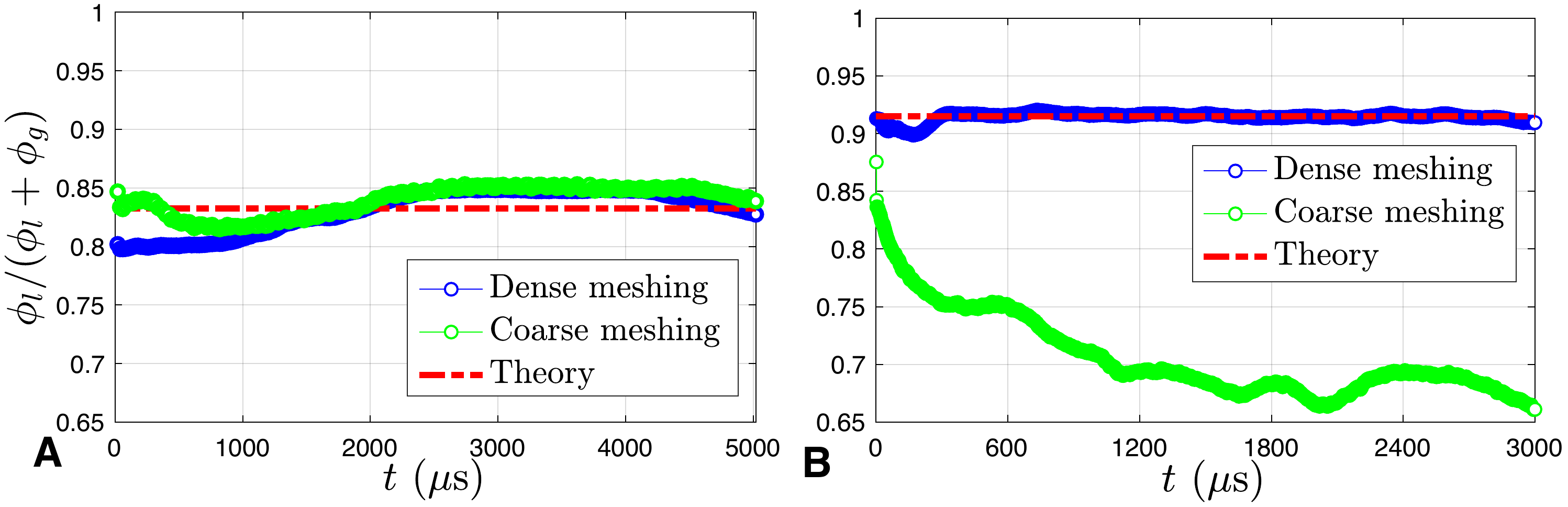}
    \caption{Comparison of the liquid ratio $\left(\phi_l / \left( \phi_l + \phi_g\right)\right)$ between the CFD computation results of the dense meshing (blue dots) and coarse meshing (green dots) and the theoretical value (red dotted line) with regards to time $t$. Note that the vertical coordinates are plotted in the range of $[0.65, 1]$ for comparison. \textbf{A.} The liquid ratio of the single bubble flow case. \textbf{B.} The liquid ratio of the multiple bubbles flow case.}
    \label{component}
\end{figure}

In multiphase flow simulations, some numerical factors (i.e., meshing, BCs, ICs, solvers) could cause losses of components, leading to inaccurate results. To validate our simulations, the liquid-gas volume ratio during the whole computation process, signifying the conservation of the components, is shown in Figure \ref{component} for both the single and multiple bubbles cases, together with the comparison with the simulations with coarse meshing. In our cases, 
if $A$ stands for the area (2D volume) of the bubbly flows, the theoretical liquid ratio (or water ratio) can be estimated by $\phi_0= A_{water} / \left(A_{water} + A_{air}\right)$. Initially,  it has $\phi_0 = 0.8324$ for the single bubble flow and $\phi_0 = 0.915$ for the multiple bubbles flow case. Based on equation (\ref{lg}), the liquid ratio of simulation can be represented as $\phi_l / \left( \phi_l + \phi_g\right)$. Figure \ref{component} shows the simulations with dense meshing generally agree with the theoretical results (blue dotted lines), with only subtle fluctuation around the theoretical value $LG$, validates the accuracy of our simulations. 
It is also found that the mass conservation is almost kept for coarse meshing simulation of the single bubble flow, yet an obvious liquid component dropping is observed for the multiple bubbles flow. Therefore, fine meshing densities are required for complex two-phase flow systems, especially for multiple bubbles flow.

\section{Deep learning algorithms \label{nn}}

In this section, we briefly introduce the basis of deep learning and NNs; then we present our approach for using DNNs and our framework \textsf{BubbleNet} to predict bubbly flows physical fields. Both the DNN and \textsf{BubbleNet} are trained on our coarsened simulation data and the bubble motion at a specific time is predicted respectively. 

\subsection{Traditional DNNs}

A neural network (NN) consists of input layers, hidden layers, and output layers, fully connected within. DNNs are NNs with multiple hidden layers that are able to approximate nonlinear operators and mapping between two Banach spaces. 

Here we apply a DNN with four sub-nets, ${\rm Net}_u,\ {\rm Net}_v,\ {\rm Net}_p,\ {\rm Net}_{\phi}$ (see Appendix), for predicting the physics fields $u,\ v,\ p,\ \phi$, respectively. Each sub-net consists of 9 layers with 30 neurons for each layer. The input quantities are the field data in the space-time domain, namely $x,\ y,\ t$. The Adam optimizer and the 'L-BFGS-B' optimization method are adopted in the training process. Each neuron is activated by the $tanh$ function. The maximum iteration for 'L-BFGS-B' optimization is 500000. 

To reduce the computation resources, the training data is obtained by coarsening the CFD results with fine meshes, which is given by interpolation: \[\left[x_{train},\ y_{train},\ t_{train}\right] = \left[x(1:\Delta^{s}:end),\ y(1:\Delta^{s}:end),\ t(1:\Delta^{t}:end) \right]\]where $\Delta^{s}$ and $\Delta^{t}$ are the spacial and temporal intervals respectively. For the single bubble case, we obtain a $2419 \times 126$ (space $\times$ time) training data on the space-time domain and $3766 \times 101$ for the multiple bubbles case. 

The coarsed data set is then normalized before training using the \textsf{mapminmax} function of \textsc{Matlab}\textregistered
. The function process matrices by mapping row minimum and maximum values to $[0,\ 1]$. If the coarsened data obtained from the simulation of bubbly flow is defined as $\mathcal{U} = (u,\ v,\ p,\ \phi)$, then the training data $\mathcal{W}$ can be obtained through normalizing $\mathcal{U}$. The function \textsf{mapminmax} normalization operates data in the full spatial-temporal domain: \begin{equation}\mathcal{W} = \frac{\mathcal{U} - \mathcal{U}_{min} }{ {\mathcal{U}_{max} - \mathcal{U}_{min}}}\label{normal}\end{equation} where $\mathcal{U}_{max}$ and $\mathcal{U}_{min}$ denote the maximum and minimum value of the coarsened simulation data in the whole space-time domain respectively. Such a process can be simplified to a normalization function $\mathbb{N}$, written as $\mathcal{W} = \mathbb{N}(\mathcal{U})$.

For the single bubble case, the DNNs are trained for 10000 iterations on the normalized data, and we aim to predict the physics fields at $t=2000\mu$s. For the single bubble case, the DNNs are trained for 200000 iterations on the normalized data, and we aim to predict the physics fields at $t=1500\mu$s.

\subsection{Semi-physics-informed neural networks}
Physics-informed neural networks (PINNs), as introduced, encode physics information into the loss function, imposing the NN to approximate the real physics equations during training. The original PINNs \cite{pinn, pinn1, pinn2} and their modified versions \cite{deepxde, deeponet} prefer to encode the whole physics governing equations, together with related BCs and ICs into the NN's loss function. Such approaches have been extensively studied and successfully applied to many physical systems, as introduced in Section \ref{intro}. However, encoding full equations into the losses might consume considerable computation resources in automatic differentiation. Furthermore, for problems like two-phase flows, there essentially exists a drastic variation of level set function $\phi$ and density at the interfaces between the two fluids, placing high demand on the accuracy of the calculation of the gradient at and around the interfaces. The inaccuracy of gradient calculation will lead to difficulties in neural network training. In traditional 
CFD, this problem is usually solved by increasing the grid density, which will inevitably raise the training time of the neural network substantially. 
As Karniadakis et al. \cite{george} claimed, if plenty of data are trained on a NN (Big data regime), PINNs can be adopted to discover new physics. In such a regime, the NNs are of ability to approximate data with good accuracy even with no physics guidance. In comparison, full physics is preferred only at the regime of small data. In the present study, we aim at predicting the bubble flow with satisfactory results based on partial physical information, to reduce the computer resource required for flow field with high gradients.

Our algorithm \textsf{BubbleNet} encodes the continuum equation of incompressible fluids and the pressure Poisson equation in the inference process, namely the semi-physics-informed neural networks, eliciting the latent function $\psi$ for predicting the velocity fields $u,\ v$:
\[u = \frac{\partial \psi}{\partial y},\ v = -\frac{\partial \psi}{\partial x}.\] 
Thus the continuum condition (divergence free of velocity), $\nabla\cdot\mathbf{u}=0$, is automatically satisfied. Meanwhile, this reduces one sub-net for 2D or axisymmetric flows, which saves considerable computation resources for initialization and training. Furthermore, the introduction of stream function avoids gradients calculation of velocity vectors in the loss function, improving the efficiency of the neutral networks.

Simultaneously, the pressure Poisson equation is also included within the losses to improve the accuracy of prediction, writes as:\[\nabla^2 p = \rho \frac{\nabla \cdot \mathbf{u}}{\Delta t} - \rho \nabla \cdot (\mathbf{u} \cdot \nabla \mathbf{u}) + \mu \nabla^2 (\nabla \cdot \mathbf{u}) \]both the processes are achieved through automatic differentiation on the output physics fields.

\subsubsection*{Time discretized normalization (TDN)}

\begin{figure}
    \centering
    \includegraphics[scale=0.7]{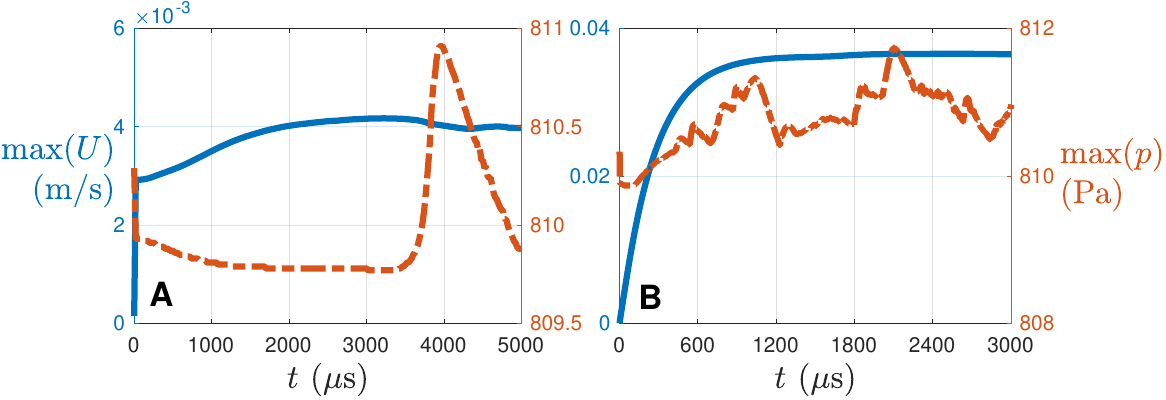}
    \caption{The variations of the maximum of velocity magnitude $U$, and pressure $p$ with time for \textbf{(A)} the single bubble flow (solid line) and \textbf{(B)} multiple bubbles flow (dashed line).}
    \label{time-evolution-of-max-values}
\end{figure}

In the present semi-PINN algorithm, the equation (\ref{normal}) is still used as normalized operator $\mathbb{N}$ acting on data set $\mathcal{U}$. However, different from the traditional method, $\mathcal{U}_{max}$ and $\mathcal{U}_{min}$ are considered as the maximum and minimum of the coarsened CFD data at each time step, which is called as time discretized normalization (TDN). The reason for this treatment lies in the significant variation of physical quantities in the flow field, as shown in Figure \ref{time-evolution-of-max-values}. In Figure \ref{time-evolution-of-max-values}, the variations of the maximum values of velocity magnitude, $U = \sqrt{u^2 + v^2}$, and pressure $p$, are depicted with respect with time. The TDN will be helpful to eliminate the inaccuracy caused by these variations.
The training data, the optimization method and the iterations are same for \textsf{BubbleNet} as in DNN.

\subsubsection*{Loss function}

In the present framework, mean squared error (MSE) is used for computing the deviation of predictions and training data in the loss function. If we use $\mathcal{W}$ to represent normalized data set, and $\mathcal{W} = (u,\ v,\ p,\ \phi)$, then the loss function $\mathcal{L}$ takes the form:
\[\mathcal{L} = \frac{1}{m} \sum_{i = 1}^m \left(\mathcal{W}_{pred(i)} - \mathcal{W}_{train(i)}\right)^2 + \frac{1}{m} \sum_{i = 1}^m \left(\nabla^2p_{(i)}\right)^2\]
where $\mathcal{W}_{pred}$ is the predictions of the NN training and $\mathcal{W}_{train}$ is the normalized training data obtained from CFD simulations. $\nabla^2p_{(i)}$ denotes pressure field in the training sets $i$. $m$ is the training data numbers.

The schematic for our proposed semi-PINN architecture \textsf{BubbleNet} is shown in Figure \ref{BubbleNetschematic}, and the details of the algorithm is described in Appendix, where we use three sub-nets: ${\rm Net}_\psi,\ {\rm Net}_p,\ {\rm Net}_\phi,$ to predict $\psi,\ p,\ \phi$ respectively, and compute the velocities through automatic differentiations from $\psi$. 

\begin{figure}
    \centering
    \includegraphics[scale=0.32]{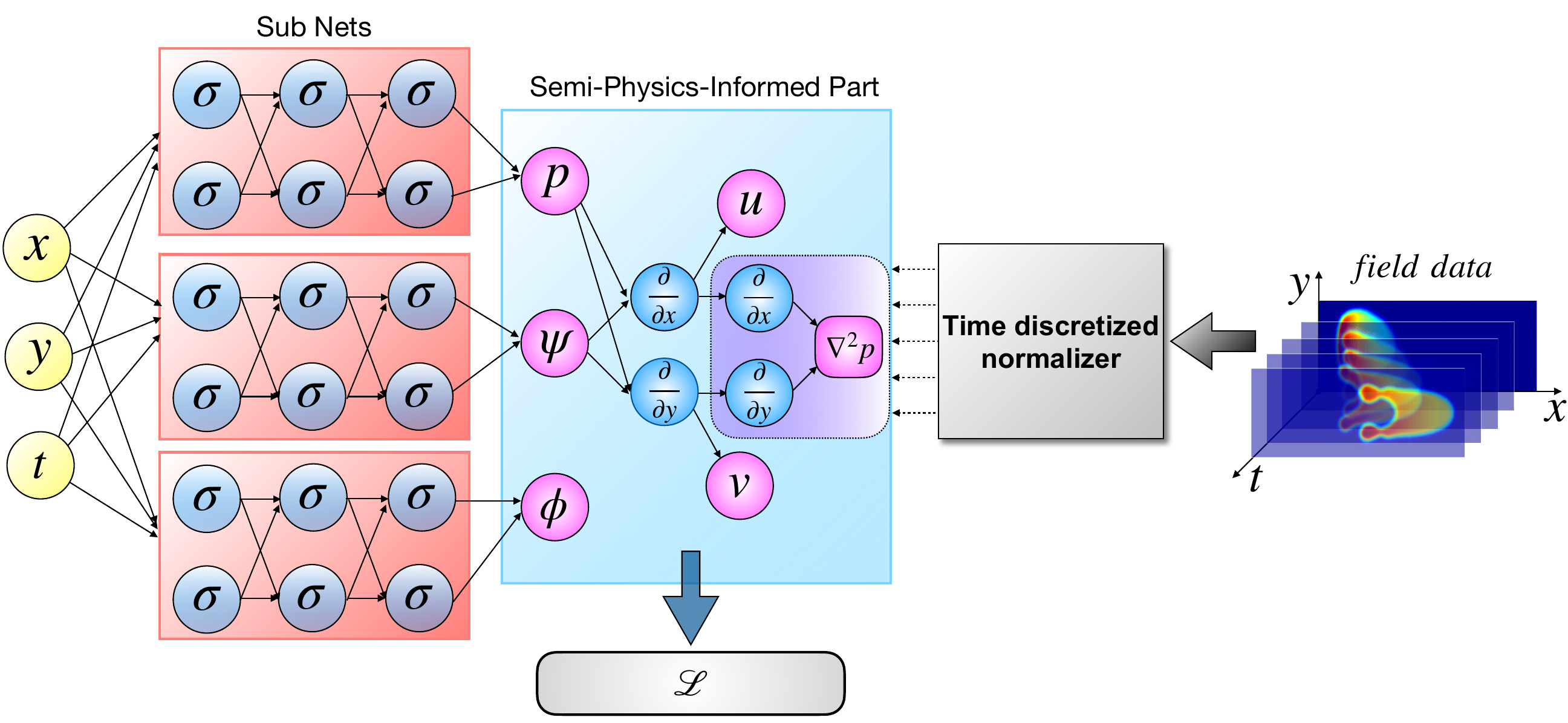}
    \caption{Schematic diagram of the deep learning framework \textsf{BubbleNet}, consisting of three sub-nets for inferring $p,\ \psi,\ \phi$, respectively, each having 9 hidden layers and 30 neurons. The semi-physics-informed part infers velocities $u,\ v$ with the automatic differentiation through the fluid continuum equation, and the pressure Poisson equation is also inserted in the loss function. The time discretized normalizer is applied to normalize the training data per time step. The Poisson equation is represented by $\mathcal{P}$ (the shaded purple part in the Diagram). The loss function consists of the residual of inferred $p,\ u,\ v,\ \phi$ and the pressure Poisson equation $\mathcal{P}$.}
    \label{BubbleNetschematic}
\end{figure}

\begin{figure}
    \centering
    \includegraphics[scale=0.72]{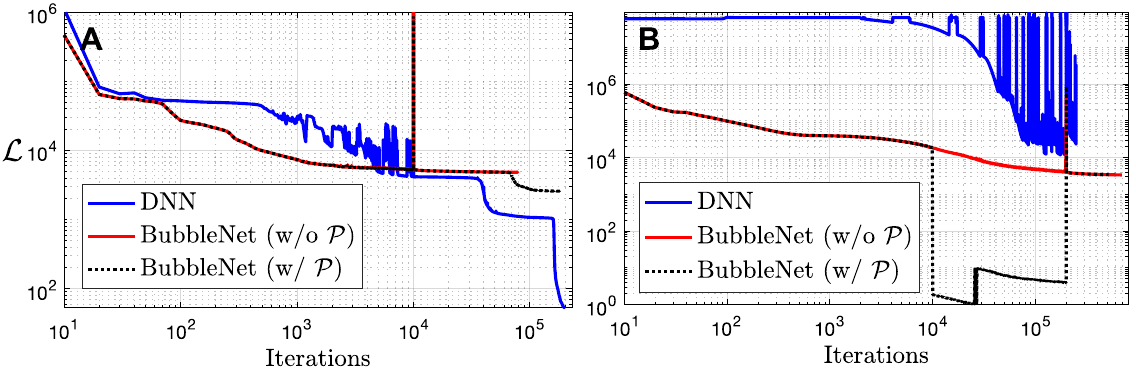}
    \caption{The variations of losses $\mathcal{L}$ in iterations of the DNNs and \textsf{BubbleNet} for \textbf{(A)} the single bubble and \textbf{(B)} multiple bubbles simulations. The blue solid line represents the losses of DNN, the red solid line for the losses of \textsf{BubbleNet} with no Poisson equation inserted, and the black dotted line denotes the losses of the \textsf{BubbleNet} with Poisson equation. }
    \label{loss_single}
\end{figure}

The variations of losses $\mathcal{L}$ in iterations of the DNNs and \textsf{BubbleNet} for both the single bubble and multiple bubbles simulations are shown in Figure \ref{loss_single}. Figure \ref{loss_single}\textbf{A} indicates traditional DNN exhibits lower losses and with longer iterations for single bubble case, whilst both \textsf{BubbleNet}(s) stop training at an earlier stage with higher losses. The higher losses may be accounted for the additional errors resulted from the physical information. Figure \ref{loss_single}\textbf{B} shows different trends: \textsf{BubbleNet} exhibits lower losses with more iterations \& training. Yet both DNNs display similar fluctuating losses (blue solid line in Figure \ref{loss_single}). Both the \textsf{BubbleNet}(s) exhibit similar magnitude and trends in losses as indicated from the red solid lines and black dotted lines. The consideration of the Poisson equation $\mathcal{P}$ has only minor effects on the variation of the losses with time. 

\section{Results of machine learning\label{results}}

\subsection{Single bubble flow\label{pred}}

\begin{figure}
    \centering
    \includegraphics[scale=0.88]{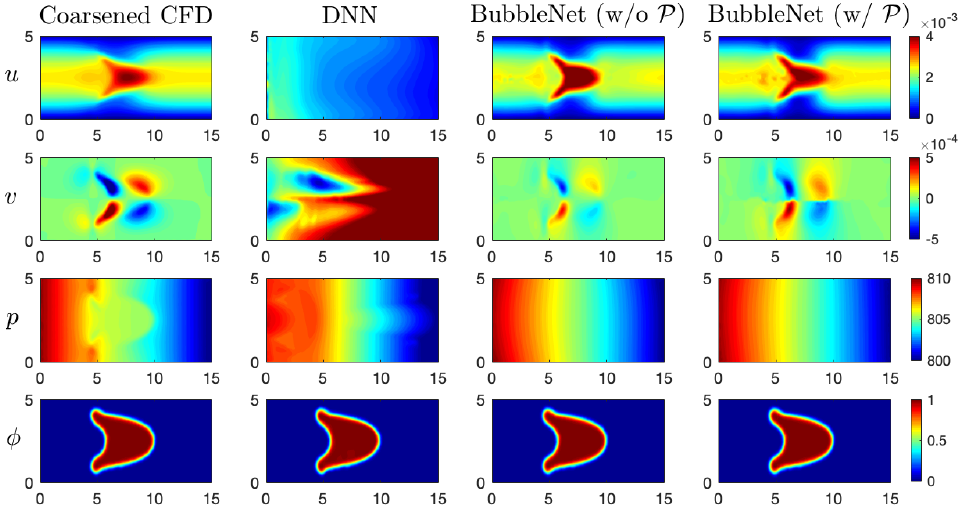}
    \caption{The comparison of the physical quantities $u,\ v,\ p,\ \phi$ obtained from the CFD simulation results (coarsened), DNN and \textsf{BubbleNet}(s). The four rows illustrate the four physical fields respectively. The first column is the coarse training data based on the CFD simulation results. The other columns illustrate the predicted physical quantities from the traditional DNN and \textsf{BubbleNet}(s).} 
    \label{single}
\end{figure}

For the single bubble flow case, both DNN and \textsf{BubbleNet}(s) are used to predict its physical variables, with comparison to the 2D CFD simulations results at $t = 2000 \mu \rm s$. The predicted physics fields $u,\ v,\ p,\ \phi$  are shown in Figure \ref{single}. \textsf{BubbleNet}(s) evidently outperforms traditional DNN on approximating the physical trends on the velocity fields. One advantage of \textsf{BubbleNet}(s) origins from the application of TDN. To eliminate the negative effects during learning induced by the significant variation of the physical quantities with time, as shown in Figure \ref{time-evolution-of-max-values}, enforcing a normalization on the time domain is obviously helpful. In comparison, normalization on the whole temporal-spatial domain results in inaccuracy in capturing the features of velocities at a specific time. 


The consideration of physical information has benefited the prediction from the neutral networks, even with only a coarse data set. The \textsf{BubbleNet} with $\mathcal{P}$ approximate quantities more accurately than \textsf{BubbleNet} w/o $\mathcal{P}$ for the two velocities fields, as compared with the third and fourth columns in Figure \ref{single}, which will be further discussed through analyzing the absolute errors of the predictions. This can be ascribed to, the Poisson equation in the losses serves as an 'inner supervision' on ${\rm Net}_p$, allowing the NN to train on the other physical variables ($u,\ v,\ \phi$) on their corresponding sub-nets more comprehensively. Including more physics is possible to further improve the performance of the networks. This result supports our idea that the semi-PINN, i.e., the combination of physical information and traditional neutral networks could be flexible in the construction of network framework, with obtaining satisfied results meeting engineering needs, especially when acquiring a huge amount of training data is impossible.

Both the DNN and \textsf{BubbleNet} (w/o $\mathcal{P}$) display good accuracy on the phase function $\phi$, and also on the overall numerical magnitudes of the pressure gradient. However, they do not successfully capture the bubble shape feature details in the pressure field, as shown in the third row in Figure \ref{single}. To overcome this defect, the Poisson equation is included in the \textsf{BubbleNet} for supervision. However, this has only a negligible influence on the results. This is because the subtle differences in pressure magnitude depicting the bubble shape are too small compared with the large pressure range ($[800, 810]$ shown in the third-row Figure \ref{single}). Such details feature missing in pressure will be discussed further. As for the prediction regarding $\phi$, the relatively large variation from 0 to 1 is easier to be detected by the NN, as shown in the fourth column of Figure \ref{single}. 

To quantitatively compare the performance of the algorithms, the absolute error $|\epsilon|$ of predictions is estimated in the present study. Suppose $\mathcal{U}_{exact}$ denotes the CFD simulation results ($u,\ v,\ p,\ \phi$), $\mathcal{U}_{pred}$ is the prediction obtained from deep learning, 
the absolute errors is defined by:
\begin{equation}
    |\epsilon_{\mathcal{U}}| = |\mathcal{U}_{pred} - \mathcal{U}_{exact}|.\label{abserr}\end{equation}
The averaged error  $\overline{|\epsilon_{\mathcal{U}}|} = \frac{\sum_{i=1}^m(|\epsilon_{\mathcal{U}}|)}{m}$ is also calculated to evaluate the algorithms.

\begin{figure}
    \centering
    \includegraphics[scale=1]{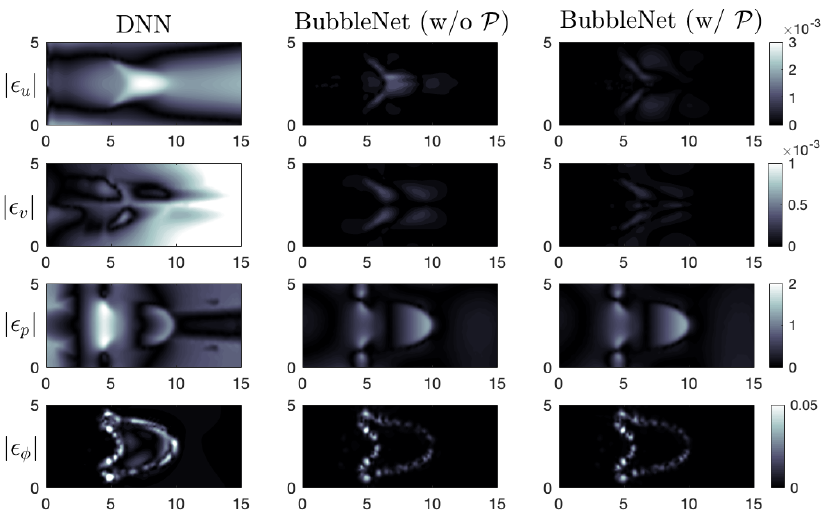}
    \caption{The comparison of the absolute errors $|\epsilon_u|$, $|\epsilon_v|$, $|\epsilon_p|$, $|\epsilon_\phi|$ between DNN and \textsf{BubbleNet} (w/ \& w/o $\mathcal{P}$) for the single bubble flow case. The four rows describe the errors of the predictions for the four physical quantities $u,\ v,\ p,\ \phi$.}
    \label{singleerror}
\end{figure}

\begin{table*}[!b]
    \centering
    \begin{tabular}{ccccc}

    \toprule
    \quad & $\overline{|\epsilon_u|}$ & $\overline{|\epsilon_v|}$ & $\overline{|\epsilon_p|}$ & $\overline{|\epsilon_\phi|}$ \\
    \midrule 
    DNN &  $9.8832\times10^{-4}$ & $5.1092\times10^{-4}$ & $0.5817$ & $0.0113$ \\
    \textsf{BubbleNet} w/o $\mathcal{P}$ & $1.2869\times10^{-4}$ & $3.9572\times10^{-5}$ & $0.2105$ & $0.0019$ \\
    \textsf{BubbleNet} w/ $\mathcal{P}$& $8.5603\times10^{-5}$ & $3.0938\times10^{-5}$ & $0.2167$ & $0.0018$ \\
    \bottomrule
    \end{tabular}
    \caption{The mean value of the absolute errors of the three deep learning frameworks for the single bubble flow case.}
    \label{errorsingle}
\end{table*}

The absolute errors corresponding to the four physical quantities $|\epsilon_u|$, $|\epsilon_v|$, $|\epsilon_p|$, $|\epsilon_\phi|$ for the single bubble flow are shown in Figure \ref{singleerror}. It can be found that both \textsf{BubbleNet}(s) display higher accuracy on the predictions of all the variables from the absolute errors, which confirms our observations in Figure \ref{single}. The \textsf{BubbleNet} w/ $\mathcal{P}$ exhibits lower errors than \textsf{BubbleNet} w/o $\mathcal{P}$ on the velocities, as indicated in the third and fourth columns of Figure \ref{singleerror}. However, significant improvement is not observed on the pressure and the phase function. To quantitatively demonstrate this result, the average absolute errors $|\overline{\epsilon_{\mathcal{U}}}|$ are computed from equation (\ref{abserr}), as shown in Table \ref{errorsingle}. It indicates that $|\overline{\epsilon_u}|$ and $|\overline{\epsilon_v}|$ are remarkably reduced for \textsf{BubbleNet} w/ $\mathcal{P}$, whereas $|\overline{\epsilon_p}|$ and $|\overline{\epsilon_\phi}|$ remain nearly unchanged for both \textsf{BubbleNet}(s) in Table \ref{errorsingle}. Together with the visualization in Figure \ref{singleerror}, it can be deduced that the two \textsf{BubbleNet}(s) displays approximately same accuracy on the two physical quantities. This implies that, the auxiliary physics-informed part $\mathcal{P}$ mainly improves the accuracy on the velocities yet not directly on the pressure field. 
It might be inferred more accurately if we increase the density of the training data set since two-order differentiation is requested in the Poisson equation. 

In summary, considering the effect of $\mathcal{P}$ on the results of the prediction, we can further expand our previous hypothesis that the 'physics-informed' part serves as inner supervision, to that the additional inner supervision may not have direct influences on its corresponding values or sub-net (${\rm Net}_p$ in our case), but is helpful to improve the overall accuracy of the predictions.



\subsection{Multiple bubble flow\label{multiple}}

\begin{figure}
    \centering
    \includegraphics[scale=1]{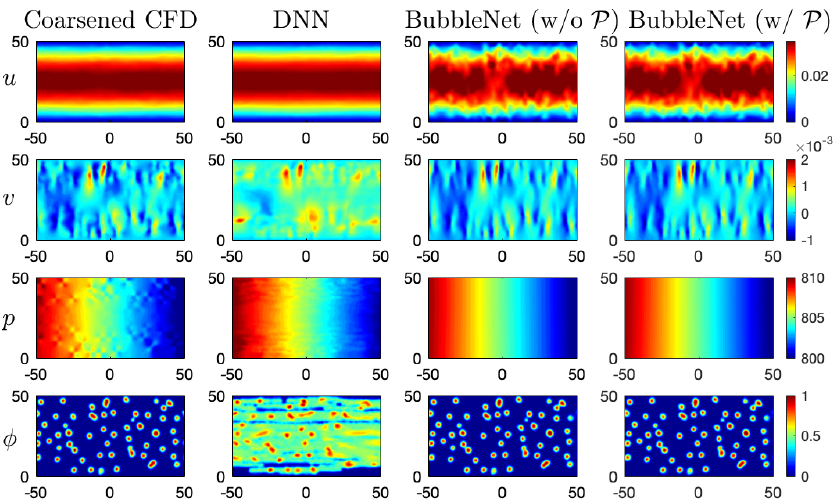}
    \caption{The comparison between the predicted physical quantities $u,\ v,\ p,\ \phi$ from the CFD simulation results (coarsened), DNN and \textsf{BubbleNet} for the multiple bubbles flow. The four rows indicates the results for velocity $u,\ v,\ p,\ \phi$, respectively. The first column indicates the coarsened training data the CFD simulation results. The other columns illustrate physical quantities from the traditional DNN and \textsf{BubbleNet}(s).}
    \label{sys}
\end{figure}

The multiple bubbles flow is more complicated to be predicted due to its significant variation in physical variables, especially for small training sets. From predictions of the deep learning frameworks shown in Figure \ref{sys}, it can be found generally \textsf{BubbleNet}(s) performs more accurately than the traditional NN, particularly on the level set function. It is not surprising that DNN fails to estimate the level set function due to its remarkable variation in the training set. 

As for the prediction of the velocity vector, \textsf{BubbleNet}(s) basically predict the general trend of its change, especially for the $y-$ component, but apparently with undesired fluctuations in its horizontal component. The introduction of the continuity condition greatly improves the prediction of the velocity in $y-$ direction. However, due to the small magnitude of velocity in this direction, the error caused by the derivation of the stream function on the coarse grid also leads to some deviation in the estimation of the velocity in $x-$ direction. In contrast, the 
DNN is better at estimating the horizontal velocity but displays a larger deviation in the vertical velocity, which reveals the advantages of the adoptions of TDN and physical equations in the inner supervision. 
Consideration of the pressure Poisson equation for supervision in the \textsf{BubbleNet} does not lead to significant improvement in the results, which can be observed in the third and fourth columns in Figure \ref{sys}. 

Quantitative error analyses for the multiple bubble flow are provided in Table \ref{errorsys}, which indicates that the inclusion of the physical equations in \textsf{BubbleNet}(s) is benefited to improve the accuracy of prediction of vertical velocity, pressure, and level set function. However, the error in the estimation of the horizontal velocity increases due to the reason we mentioned above. Moreover, different from the single bubble case, there are only negligible differences between the two \textsf{BubbleNet}(s). This implies that to get results that meet engineering needs, we may only need to consider part of the physical information in the neural network, instead of full Navier-Stokes equations. This will profoundly reduce the computational resources requested for neural network training. Further quantitative analysis of the effect of the quantities of required physics information for the training of PINNs might be a promising topic.

\begin{table*}[!b]
    \centering
    \begin{tabular}{ccccc}

    \toprule
    \quad & $\overline{|\epsilon_u|}$ & $\overline{|\epsilon_v|}$ & $\overline{|\epsilon_p|}$ & $\overline{|\epsilon_\phi|}$ \\
    \midrule 
    DNN &  $3.0235\times10^{-4}$ & $3.8237\times10^{-4}$ & $0.5725$ & $0.3492$ \\
    \textsf{BubbleNet} w/o $\mathcal{P}$ & $0.0015$ & $7.6516\times10^{-5}$ & $0.2525$ & $0.0061$ \\
    \textsf{BubbleNet} w/ $\mathcal{P}$& $0.0015$ & $7.7402\times10^{-5}$ & $0.2481$ & $0.0075$ \\
    \bottomrule
    \end{tabular}
    \caption{The mean value of the absolute errors fields of the three deep learning frameworks for the multiple bubbles flow case.}
    \label{errorsys}
\end{table*}

The level set function is crucial to identify the structures and dynamics of bubbles. It is impressive that \textsf{BubbleNet}(s) present remarkable accuracy in the prediction of $\phi$ for the complex flow, in which $|\epsilon_\phi|$ reduces by several orders after the continuous equation is imposed. It could be difficult for traditional NN due to the large gradient on the surface and the small training data set, especially on coarse grids. This manifests the advantage of the present algorithm in its simplicity and accuracy. 

Both the DNN and \textsf{BubbleNet}(s) succeed in approximating the horizontal variation of the pressure field, and the results obtained by the \textsf{BubbleNet}(s) is more accurate, as shown in the third row in Figure (\ref{sys}) and Table \ref{errorsys}. However, both algorithms miss the subtle variations in pressure field caused by the bubble(s) movement, as in Figures \ref{single}, \ref{sys}, even the pressure Poisson equation is utilized for supervision. 

\begin{figure}
    \centering
    \includegraphics[scale=0.2]{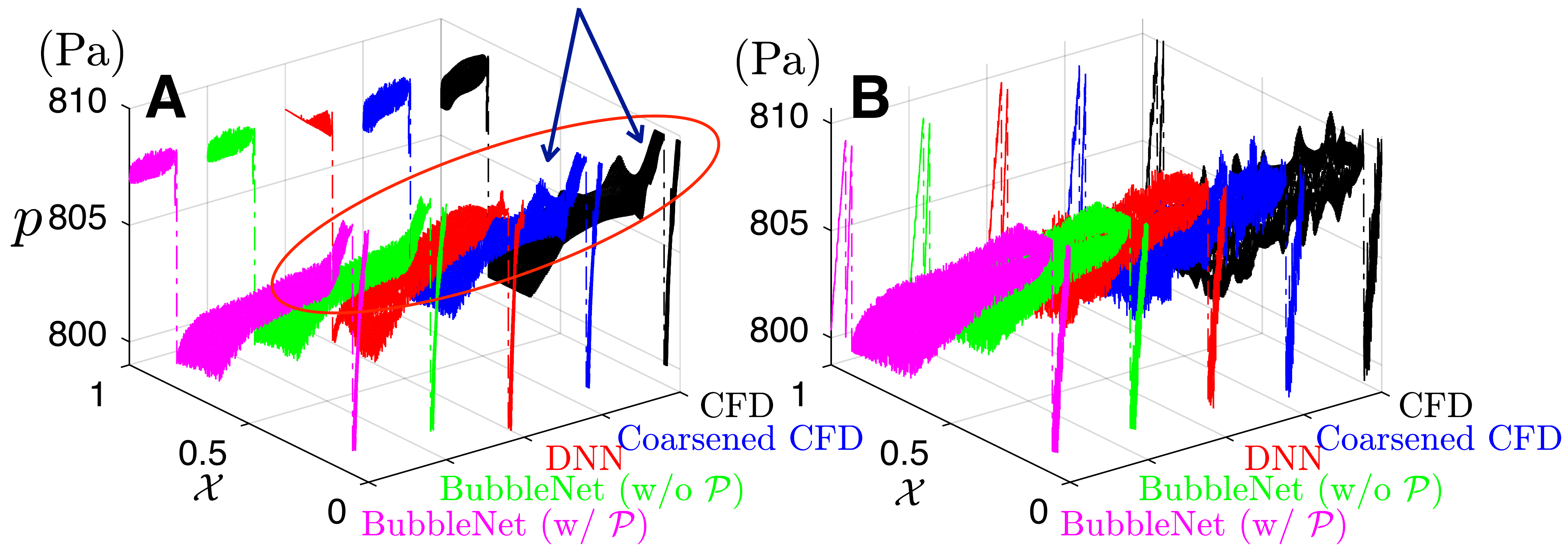}
    \caption{The comparison of the pressure distribution for both the two bubbly flow cases. The black dotted lines illustrate the original CFD numerical results, with dense meshing. The blue dotted lines represent for the coarsened CFD data used for neural network's training. The red dotted lines stand for the results of the traditional deep neural network's. The blue dotted lines and the pink dotted lines correspond to the \textsf{BubbleNet}'s predictions without and with Poisson equation respectively. \textbf{A.} Single bubble flow. \textbf{B.} Multiple bubbles flow. }
    \label{coarse-prediction-comparison}
\end{figure}

In order to find the reason for the loss of pressure information, the data distribution of $p$ at a targeted time is depicted in Figure \ref{coarse-prediction-comparison}, where the original data of CFD (with dense meshing), the coarsened CFD data for NN training (training), DNN and \textsf{BubbleNet}(s) predictions are plotted on the normalized space domain $\mathcal{X}$ for the two bubbly flow cases. Here $\mathcal{X}$ represents for the normalized form of the space field $\mathbb{X} = \left[ (x_1,\ y_1),(x_2,\ y_2),...,(x_n,\ y_n)\right]$, and $n$ is the number of meshing elements. In the right-top corner of Figure \ref{coarse-prediction-comparison} \textbf{A} (marked with the red circle), a crater-like shape (pointed out by blue arrows), which describes the structure of the single bubble, can be found for the two CFD data sets, whereas it is absent in all the predicting results. The present DNN and \textsf{BubbleNet}(s) fail to capture this subtle feature containing important physical information in pressure. Similar phenomenon is also observed in Figure \ref{coarse-prediction-comparison} \textbf{B} for multiple bubble flow. The data coarsening might be partly accounted for this 'feature losing', yet increasing data density will inevitably lead to a rise in computational resources. Another more promising strategy might be improving the network structure in the present study, i.e., using U-Net, GAN, ConvLSTM, etc., which have been successfully applied to turbulence investigations \cite{turbexp}. 

\begin{figure}[t]
    \centering
    \includegraphics[scale=0.85]{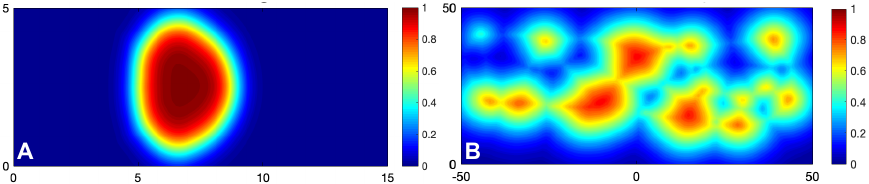}
    \caption{The simulation results for the two bubbly flow cases with the coarse meshing, for the comparison with the deep learning results presented in Figures \ref{single} \& \ref{sys}. \textbf{A}. Single bubble flow. \textbf{B}. Multiple bubbles flow.}
    \label{cfd-dnn-bubblenet-coarse-comparison}
\end{figure}

Finally, as an additional evaluation of the present algorithms, the CFD simulations for single and multiple bubbles flows are conducted on the coarse meshing we use for training, as shown in Figure \ref{cfd-dnn-bubblenet-coarse-comparison}.
Apparently, these numerical results are inaccurate in calculating the liquid-gas interface and the bubble deformations, see Figures \ref{single}, \ref{sys} for references. This further validates the effectiveness of the semi-PINN presented in this study.


\section{Conclusions \label{conclusion}}
In the present study, both traditional DNN and semi-PINN framework \textsf{BubbleNet} are applied for inferring and predicting the physics information of bubbly flow. To obtain training data set, two-dimensional simulations by the conventional CFD software are conducted for the single and multiple bubbles' flows. We briefly analyze their flow fields and find the liquid mass ratio remains nearly constant during the whole computation, which verifies the reliability of our computations.

The deep learning framework \textsf{BubbleNet} proposed in this investigation consists of the DNN with three sub-nets for predicting different physics fields (specifically $\psi,\ p,\ \phi$), the semi-physics-informed part with the fluid continuum condition, and the pressure Poisson equation $\mathcal{P}$ encoded, and the time discretized normalizer (TDN), which is adopted to normalize physical variables per time step before training. The purpose of constructing this architecture is to avoid high order differentiation if the full hydrodynamic equations are encoded into the loss function, which will lead to difficulty in reducing the differential error due to the large gradient near the liquid-gas interface in bubbly flows. The \textsf{BubbleNet} framework with and without $\mathcal{P}$ are considered separately to reveal how physics-information works in neutral networks. Considering the high resolution of flow fields sometimes can hardly be obtained, the training data set is intentionally coarsened through interpolation from the original CFD simulation results to reduce the computational consumption of machine learning.



The effectiveness of the \textsf{BubbleNet}(s) is demonstrated from training the coarsened data obtained for the two bubbly flow cases. Results indicate the \textsf{BubbleNet} is of ability to predict the physics more accurately than the traditional DNNs, in which the absolute errors of the physical quantities $|\epsilon|$ decrease profoundly, especially for multiple bubble flow. TDN is also helpful to improve the accuracy of the algorithm, which indicates that a proper characteristic scale is crucial in machine learning, especially for small data sets. The inclusion of the Poisson equation has a limited effect on reducing absolute errors of machine learning. 

In summary, although the deep neutral network encoding full hydrodynamic equation might be more accurate in prediction, the present \textsf{BubbleNet}, which is essentially an engineering-orientated semi-PINN, has the advantages in simplicity, computational efficiency, and flexibility. This raises an intriguing question that deserves to be pursued in the future, namely that we can optimize the network performance by selectively introducing physical information into the neural network.



\section*{Acknowledgments}
This research is supported by the Natural Science Foundation of China (No. 11772183, 11832017 and 12172209).

\section*{Data availability}
The data and code used in this paper can be downloaded at \url{https://github.com/hanfengzhai/BubbleNet}. The simulation programs, data, and other related files can be accessed through \url{https://doi.org/10.5281/zenodo.4632466}. Details of this project can be viewed at \url{https://hanfengzhai.net/BubbleNet}.

\clearpage
\section*{Appendix. Algorithms for DNN \& BubbleNet}

The algorithms for DNN and \textsf{BubbleNet} used in the paper are here attached as in \textbf{Algorithm 1} and \textbf{Algorithm 2}. In \textbf{Algorithm 1} we use the \textsf{MaxMinScaler} to represent usual normalization method. For the single bubble case, \textit{\#Iterations} $=10^4$, and \textit{\#Iterations} $=2 \times 10^5$ for the multiple bubbles case. For the single bubble case, $t_{pred} =2000\mu\rm s$; for the multiple bubbles case, $t_{pred} =1500\mu\rm s$. The four sub nets $ {\rm Net}_u,{\rm Net}_v,{\rm Net}_p,{\rm Net}_\phi$ are executed on four separate \textsf{def} functions. Note that the codes are run on \texttt{tensorflow 1.15.0}.

\begin{center}
    \includegraphics[scale=1]{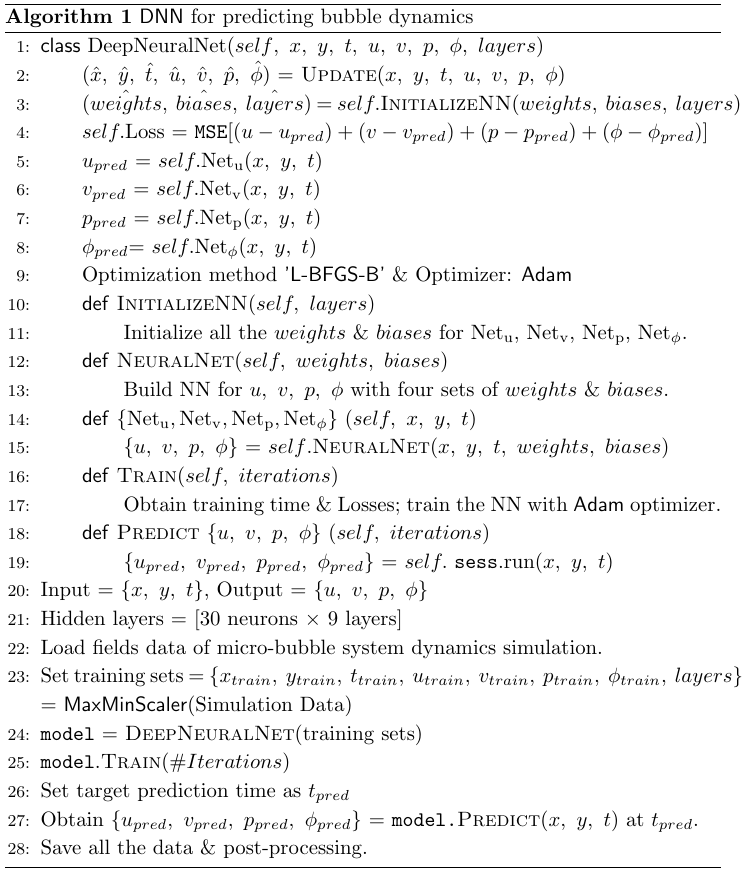}
\end{center}
\begin{center}
    \includegraphics[scale=1]{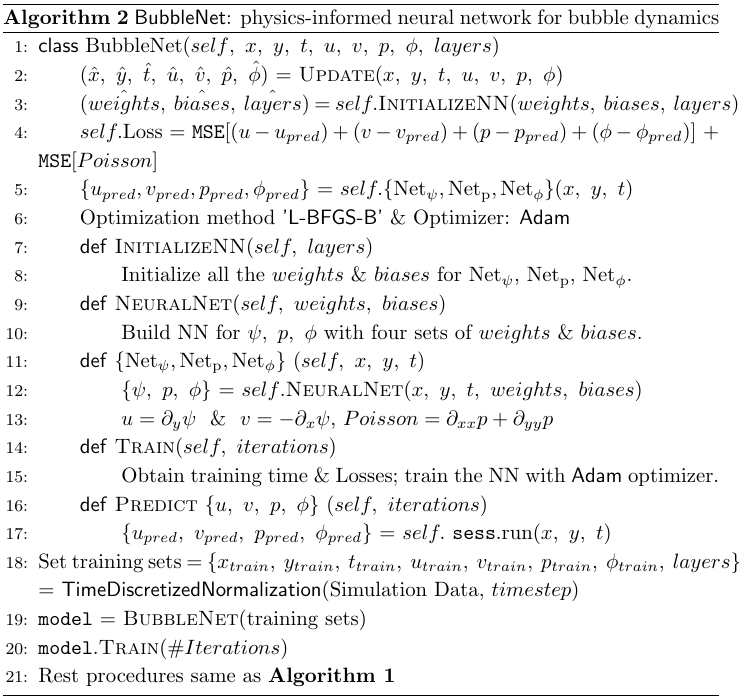}
\end{center}
\end{document}